\documentclass[aps,prb,reprint]{revtex4-2}
\usepackage{amsmath,amssymb,amsthm,hyperref,graphicx,physics,subcaption,float,dsfont}
\hypersetup{hidelinks}
\usepackage[capitalise]{cleveref}
\hypersetup{colorlinks=true, linkcolor=blue, citecolor=red, urlcolor=blue}
\newcommand{\ii}{\mathrm{i}}
\newcommand{\kB}{ } 
\newcommand{\pF}{p_{\mathrm{F}}}
\newcommand{\nF}{n_{\mathrm{F}}}
\newcommand{\pfnu}{p_{\mathrm{F},\nu}}
\newcommand{\wD}{\omega_{\mathrm{D}}}
\newcommand{\vF}{v_{\mathrm{F}}}
\newcommand{\F}{\mathcal F}
\newcommand{\G}{\mathcal G}
\newcommand{\N}{\mathcal N}
\renewcommand{\Re}{\operatorname{Re}}

\newcommand{\rev}[1]{{\color{black} #1}}
\newcommand{\revtwo}[1]{{\color{black} #1}}

\begin{document}
	\title{\rev{Field-induced first order quantum phase transition of the superconducting order parameter in Ising superconductors with Rashba spin-orbit coupling}}
	\author{J. S. Harms}
	\email{joren.harms@uni-konstanz.de}
	\affiliation{Fachbereich Physik, Universit\"at Konstanz, D-78457 Konstanz, Germany}
	\author{M. Hein}
	\affiliation{Fachbereich Physik, Universit\"at Konstanz, D-78457 Konstanz, Germany}
	\author{W. Belzig}
	\affiliation{Fachbereich Physik, Universit\"at Konstanz, D-78457 Konstanz, Germany}
	\date{\today}
	\begin{abstract}
			\rev{Ising superconductors have attracted significant attention recently due to their resilience against magnetic fields far beyond the Pauli-paramagnetic limit as well as their broader role as a platform for exploring the interplay of spin-orbit coupling, valley degrees of freedom and superconducting pairing in two-dimensional materials.
			Their protection against external magnetic field relies on strong Ising spin-orbit coupling, which originates from in-plane inversion symmetry breaking.
			Due to the heavy atom nature of Ising SCs, a smaller but sizable Rashba SOC can arise from gating or interfacial effects.
			Here, we study Rashba spin-orbit coupling within a two-valley model of Ising superconductors with an attractive $s$-wave interaction.
			We show that Rashba SOC induces a first order quantum phase transition in Ising superconductors.  
            At low temperatures, this means the superconducting order parameter vanishes abruptly in the vicinity of the quantum critical point.
            This effect, however, disappears at high temperatures.
			Our findings demonstrate that the low- and high temperature physics of Ising SCs are quantitatively and qualitatively different in the two-valley model.}
		\end{abstract}
	\maketitle
	\section{Introduction}
	Ising superconductors (SC) are doped two-dimensional transition metal dichalcogenides (TMDs), which are characterized by strong resilience against in-plane magnetic fields beyond the Pauli-paramagnetic limit.
	As a platform, TMDs are under active interest due to their direct band gap atomic scale thickness and strong spin-orbit coupling (SOC)~\cite{manzeli_2d_2017}.
	Like graphene, TMDs have a honeycomb lattice and hence two valleys.
	Unlike  graphene, however, TMDs exhibit strong SOC because of their heavy atoms and in-plane inversion symmetry breaking.
	These features make them particularly interesting for potential applications in optoelectronics, valleytronics, and spintronics~\cite{xiao_coupled_2012,mak_control_2012}.
	When sufficiently doped, TMDs can become superconducting (SC)~\cite{kuzmanovic_tunneling_2022,de_la_barrera_tuning_2018,hamill_two-fold_2021,simon_transition-metal-dichalcogenide_2024,ilic_enhancement_2017,mockli_ising_2020,ramires_tailoring_2018,mockli_ising_2020,mockli_magnetic-field_2019,costanzo_tunnelling_2018}.
	Due to their large SOC, Cooper pairs are Ising coupled, making them resistant to large external magnetic fields~\cite{kuzmanovic_tunneling_2022,de_la_barrera_tuning_2018,hamill_two-fold_2021,simon_transition-metal-dichalcogenide_2024,ilic_enhancement_2017,mockli_ising_2020,ramires_tailoring_2018,mockli_ising_2020,mockli_magnetic-field_2019,costanzo_tunnelling_2018}.

	Besides the Ising SOC, one expects a Rashba SOC to originate from gating or \rev{the interface} in experiment---due to the heavy atoms of the Ising SC.
	Previously, it has been shown that the presence of Rashba SOC can lead to topological superconducting phases~\cite{yuan_possible_2014,meidan_josephson_junction_2024,schaffer_crystalline_nodal_topology_2020}.
	Experimentally, the strength of Rashba SOC has been claimed to be of the order of the SC order parameter~\cite{lu_evidence_2015,lu_full_2018}, making Rashba SOC possibly important for understanding the phase diagram of Ising SCs.

	\revtwo{In this article, we show that Rashba SOC introduces a first order quantum phase transition.
    This means Rashba SOC sets a critical magnetic field---in the $s$-wave two valley model of Ising SCs---beyond which the superconducting order parameter (SC OP) vanishes abruptly in a magnetic field interval proportional to temperature $T$.
    This transitions becomes a first order phase transition at zero temperature, rendering this a first-order quantum phase transition.
	While the impact of Rashba SOC is usually subtle, e.g., in the superconducting diode effect~\cite{yuan_topological_2021,ilic_theory_2022,khodas_diode_effect_Ising_2025}, the effect of the QPT is rather pronounced.
	Our results show a qualitative and quantitative difference between the low- and high temperature behavior of ISCs.
	Furthermore, our results may lead to new ways to determine the strength of the Rashba SOC in Ising SCs, which is currently a challenging task.}

	\section{The model}
\begin{figure}[t]
			\includegraphics{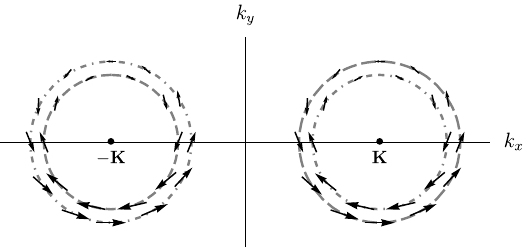}
			\caption{Graphical depiction of the in-plane component of the spin-splitting at the Fermi surface due to Rashba SOC and an external magnetic field in the $x$-direction.
				The out of plane spin splitting is given by the Ising SOC.}
			\label{fig:fermi-surface}
		\end{figure}

	The low energy physics of doped transition metal dichalcogenides is described by two valleys around the $\mathrm K$ and $\mathrm K'=-\mathrm K$ points and a gamma pocket $\Gamma$.
	In the proceeding, we focus on the physics of the $\pm\mathrm K$ points and ignore the physics of the Gamma pocket.
	The general Hamiltonian near the $\pm\mathrm K$~\cite{xiao_coupled_2012,kormanyos_k_2015,mockli_ising_2020,ilic_enhancement_2017} points in the $(a_{\mathbf p,\uparrow},a_{\mathbf p,\downarrow})$ basis reads
	\begin{equation*}\label{eq:Hamiltonian_initial}
		\hat H_0(\mathbf p+\eta\mathbf K)=
		\left(\frac{\mathbf p^2}{2m^*}-\mu\right)\sigma_0
		+\eta\beta\sigma_z
		+\alpha(\hat z\times\mathbf p)\cdot\boldsymbol\sigma
		+\mathbf H\cdot \boldsymbol\sigma
	\end{equation*}
	with $\eta=\pm 1$ the valley number, $\boldsymbol\sigma$ the matrix of Pauli matrices and $\mathbf H$ the external magnetic field.
	The SOC coupling consists of an Ising SOC $\eta\beta\hat z$ and a Rashba SOC $\alpha(\hat z\times\mathbf p)$, see~\cref{fig:fermi-surface}.
	We consider the interactions to be an attractive four point interaction towards $s$-wave superconductivity.
	Consequently, the mean-field Bogoliubov-de-Gennes (BdG) Hamiltonian in the basis $ \big(a_{\mathbf{k},\uparrow},a_{\mathbf{k},\downarrow},a_{\mathbf{-k},\uparrow}^\dagger,a_{\mathbf{-k},\downarrow}^\dagger\big)$ becomes
	\begin{equation}\label{eq:BdG-Hamiltonian}
		\hat H_\mathrm{BdG}=
		\begin{pmatrix}
			\hat H_0(\mathbf{k})&\ii \sigma_y \Delta\\
			-\ii \sigma_y \Delta^*&-\hat H_0^*(\mathbf{-k})
		\end{pmatrix},
	\end{equation}
	with $\mathbf k=\mathbf p+\eta \mathbf K$ and $^*$ denotes complex conjugation.
	Moreover, $\Delta$ is the SC singlet order parameter (OP).
	From the BdG Hamiltonian, the Gorkov Green's function is defined as
	$
	\check{{\mathcal G}}_{\ii \omega_n}^{-1}(\mathbf k)
	=
	\ii  \omega_n-\hat H_\mathrm{BdG}.
	$

	To simplify calculations, we diagonalize the normal state in spin space using the unitary transformation
		$
		\check{\mathcal{U}}_\mathbf k=\hat{U}_\mathbf k\oplus\hat U^*_\mathbf{-k},
		$
		with
		$
		\hat U_\mathbf k
		=
		e^{-{\ii }\phi_\mathbf k\sigma_z/2}e^{-\ii \theta_\mathbf k\sigma_y/2}.
		$
	The angles used above are defined via
	$
	[\sin(\theta_\mathbf k)\cos(\phi_\mathbf k),\sin(\theta_\mathbf k)\sin(\phi_\mathbf k),\cos(\theta_\mathbf k)]
	=
	{(\eta\beta\hat z+\alpha(\hat z\times\mathbf p)+\mathbf H)}/
	{\|\eta\beta\hat z+\alpha(\hat z\times\mathbf p)+\mathbf H\|}$. \rev{Note that this is a local basis transformation defining different quantization axes in both valleys.}
	The SC OP transforms equivalently and reads
	$
	\Delta\ii \sigma_y
	\rightarrow
	\Delta
	\hat{U}_\mathbf{k}^\dagger\ii \sigma_y\hat{U}^*_\mathbf{-k}
	=
	\Delta
	(\bar\beta\sigma_0+\bar H\ii \sigma_y)
	+\mathcal O(\alpha^2),
	$
	with $\bar\beta={\beta}/{\sqrt{H^2+\beta^2}}$ and $\bar H=H/\sqrt{H^2+\beta^2}$.
	By grouping together spin-up and spin-down sectors in the BdG Hamiltonian~\eqref{eq:BdG-Hamiltonian}, the inverse Green's function splits into a BCS-like sector with a rescaled SC OP $\Delta\rightarrow\bar\beta\Delta$ and an off-diagonal sector governed by the external magnetic field strength $\bar H$
	\begin{align*}\label{eq:inverse-greens-function}
		\check{\mathcal{G}}_{\ii \omega_n}^{-1}(\mathbf k)
		=\nonumber
		\ii \omega_n\mathbb{I}
		&-
		\bigoplus_{\nu=\pm}
		\begin{pmatrix}
			\epsilon_\nu(\mathbf{k})&\bar\beta\Delta\\
			\bar\beta^*\Delta^*&-\epsilon_\nu(\mathbf{-k})
		\end{pmatrix}
		-\ii \sigma_y\otimes
		\delta \check H,
	\end{align*}
		where we introduced
		$
		\delta \check H
		=
		\begin{pmatrix}
			0&\bar H\Delta\\
			-\bar H^*\Delta^*&0
		\end{pmatrix}
		$
		for notational convenience.
        \rev{Since $|\Delta|/\sqrt{H^2+\beta^2} \lesssim 10^{-1}$ in Ising SCs, the off-diagonal term $\mathrm{i}\sigma_y \otimes \delta\check H$ appearing in the inverse Green's function constitutes a small correction and may therefore be treated perturbatively.~\cite{de_la_barrera_tuning_2018,falson_type-ii_2020,hamill_two-fold_2021,xi_ising_2016,kuzmanovic_tunneling_2022,lu_full_2018,lu_evidence_2015}}
		The lowest order Gorkov Green's function has a BCS form and reads
		\begin{align}\nonumber
			\check{\mathcal{G}}^0_{\ii \omega_n,\nu}
			\simeq&
			\frac{
				\begin{pmatrix}
					\ii \omega_n+\epsilon_\nu(\mathbf{-k})&\bar\beta\Delta\\
					\bar\beta^*\Delta^*&\ii \omega_n-\epsilon_\nu(\mathbf{k})\\
				\end{pmatrix}
			}
			{
				(\ii \omega_n-\epsilon_\nu(\mathbf{k}))
				(\ii \omega_n+\epsilon_\nu(\mathbf{-k}))-|\bar\beta\Delta|^2
			},
		\end{align}
		with $\epsilon_\nu(\pm\mathbf k)=\mathbf p^2/2m^*+\nu\sqrt{H^2+\beta^2}\pm\alpha\nu (\mathbf {\bar {H}}\times\hat z)\cdot\mathbf p$.
		We obtain the contribution from the external magnetic field by expanding the Gorkov Green's function up to first order in $\ii \sigma_y\otimes\delta\check H$,
		\begin{equation}
			\begin{aligned}
				\label{eq:1st-order-greens-function}
				\check{\mathcal{G}}_{\ii \omega_n}
				\simeq&
				\begin{pmatrix}
					\check{\mathcal{G}}^0_{\ii \omega_n,+}&-\check{\mathcal{G}}^0_{\ii \omega_n,+}\delta H  \check{\mathcal{G}}^0_{\ii \omega_n,-}\\
					-\check{\mathcal{G}}^0_{\ii \omega_n,-}\delta H^\dagger\check{\mathcal{G}}^0_{\ii \omega_n,+}&\check{\mathcal{G}}^0_{\ii \omega_n,-}
				\end{pmatrix},
			\end{aligned}
		\end{equation}
		which gives the Green's function up to the desired accuracy.
		Since our goal is to determine the self-consistency equation, we only need the anomalous Green's function.
		Within the quasi-classical approximation, the diagonal part of the anomalous Green's function in~\cref{eq:1st-order-greens-function} reads
		\begin{align*}
			&\frac{f_{\ii \omega_n,\nu,\nu}^0(\eta,\phi)}{\bar\beta\Delta}
			=
			\frac
			{2\pi \ii \N}
			{\ii \sqrt{\left(\omega_n+\ii \big[\alpha\nu (\mathbf{\bar H}\times\hat{z})\cdot\mathbf p_{F}\big]\right)^2+|\bar\beta\Delta|^2}},
		\end{align*}
		while the off-diagonal, magnetic field dependent, contributions give
		\begin{align*}
			\frac{f_{\ii \omega_n,+,-}^0(\eta,\phi)}{\bar H\Delta}
			=&
			\frac
			{2\pi\ii \N\mathrm{sign}(\omega_n)}
			{\ii \omega_n-\sqrt{H^2+\beta^2}},
			\\
			\frac{f_{\ii \omega_n,-,+}^0(\eta,\phi)}{-\bar H\Delta}
			=&
			\frac
			{2\pi\ii \N\mathrm{sign}(\omega_n)}
			{\ii \omega_n+\sqrt{H^2+\beta^2}}.
		\end{align*}
		We defined $\mathcal N$ to be the electron density of states in the normal state \rev{and $\phi$ the angle on the Fermi surface.}

		From here, we are in the position to write down the self-consistency equation.
		In the spin-diagonal basis,
		this reads
		\[
		\frac{\Delta}{\lambda}
		=
		\frac{\kB T}{2}\sum_n\int\frac{\mathrm d\phi}{2\pi}
		\mathrm{Tr}\left\{f_{\ii \omega_n}\left(\mathcal U_\mathbf{-k}^T\ii \sigma_y\mathcal U_\mathbf k\right)\right\}.
		\]
		Here, $\lambda$ denotes the magnitude of the attractive four point interaction, which relates directly to the SC OP at zero temperature $1/\mathcal N \lambda=\log(2\wD/|\Delta_0|)$, with $\wD$ the Debye frequency.
		Accordingly, the self-consistency equation for the SC OP becomes
		\begin{align}\label{eq:self-consistency-equation}
			\frac{1}{\mathcal N\lambda}
			=
			\pi \kB T\sum_{n,\nu}\int\frac{\mathrm d\phi}{2\pi}\;
			\Bigg[
			\frac{\bar H^2\mathrm{sign}(\omega_n)}{\omega_n+\ii \nu\sqrt{H^2+\beta^2}}&
			\\\nonumber+
			\frac{\bar\beta^2}{\sqrt{\left(\omega_n+\ii \big[\alpha\nu (\mathbf{\bar H}\times\hat{z})\cdot\mathbf p_{F}\big]\right)^2+|\bar\beta\Delta|^2}}\Bigg]&.
		\end{align}

		In the following, we discuss the impact of Rashba SOC on the magnetic field dependence of the SC OP at low temperatures.
		\rev{In principle, Rashba SOC has two main consequences in Ising SCs.
	    First, } Rashba SOC alters the thermodynamic ground state to a helical state $\Delta\rightarrow\Delta_0e^{\ii  q x}$~\cite{kaur_helical_2005,yuan_topological_2021}.
		This \rev{contribution} is however negligible with respect to the magnitude of the SC OP since $\vF q$ is at most of the order $\mathcal{O}(\alpha \pF /\beta)$ and the correction of the helical phase to $|\Delta|/\Delta_0$ is at most of the order \rev{$\mathcal O(\alpha^2\pF^2/\beta^2)\lesssim 10^{-2}$}.
		However, the dominant influence of Rashba SOC on the magnitude of the SC OP is that it induces a first-order quantum phase transition.
        This means that the SC OP vanishes in an interval proportional to temperature around the critical magnetic field set by Rashba SOC, as we will show next.

        \begin{figure*}[t]
            \centering
            \begin{subfigure}[b]{.49\textwidth}
                \centering
                \includegraphics[scale=1]{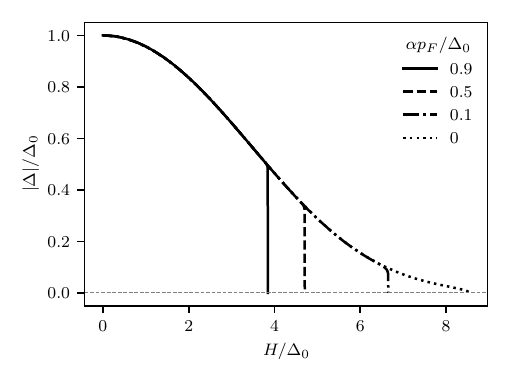}
            \end{subfigure}
            \hfill
            \begin{subfigure}[b]{.49\textwidth}
                \centering
                \includegraphics[scale=1]{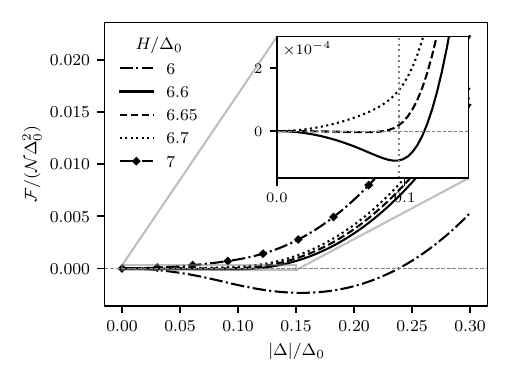}
            \end{subfigure}
			\caption{(left) Numerical solution of the self-consistency equation~\cref{eq:self-consistency-equation} for $\bar\beta|\Delta|$ as a function of the external magnetic field $H$ for different values of Rashba SOC $\alpha\pF$ at temperature $T/T_c=10^{-2}$ and Ising SOC strength $\beta=7\Delta_0$. (right) The normalized free energy with respect to the normalized gap for the same values of $\beta,T_c$ as well as for $\alpha\pF=0.1\Delta_0$ and different values of $H$. The dotted vertical line in the inset marks the minimum of the free energy at the field $H_\alpha$.}
            \label{fig:free-energy-main}
		\end{figure*}
		\section{Field induced first order QPT induced by Rashba SOC}
		In the following, we focus on low temperatures $\kB T\ll\alpha \pF\lesssim \Delta_0\ll\sqrt{H^2+\beta^2}$.
		Translated to the critical temperature, this means $T/T_c\lesssim10^{-2}$.
		Since \rev{$T_c\lesssim10\mathrm K$} in Ising SCs, this corresponds to temperatures of the order \rev{of $100\mathrm{mK}$}~\cite{kuzmanovic_tunneling_2022,hamill_two-fold_2021,li_recent_2021,xi_ising_2016}.
		Such temperatures are typically accessible in experiment~\cite{kuzmanovic_tunneling_2022}.
		Furthermore, we consider the energy scale of Rashba SOC to be at most of the order of the energy scale of the SC OP at zero temperature.
		As detailed in the supplementary material, we perform the summation over Matsubara frequencies in the first term of~\cref{eq:self-consistency-equation}, which is known to give the digamma function~\cite{kopnin_theory_2009} and approximate the second term for low temperatures.
		As a result, the self-consistency equation~\eqref{eq:self-consistency-equation} for low temperatures
		reads
		\begin{align}\label{eq:self-consistency-equation-final}
			&\log\left(\frac{|\bar\beta\Delta|}{\Delta_0}\right)
			+\frac{H^2}{\beta^2}
			\left[
			\log\left({2\sqrt{H^2+\beta^2}}/{\Delta_0}\right)
			\right]
			\\\nonumber
			&=-\frac{\kB T}{\sqrt{|\bar\beta\Delta|\alpha \pF\bar H}}
			\log\left(1+\exp\left[\frac{\alpha \pF\bar H-|\bar\beta\Delta|}{\kB T}\right]\right).
		\end{align}
		The left-hand side of~\cref{eq:self-consistency-equation-final} describes the magnetic field dependence of the Ising SC in the absence of Rashba SOC.
		The right-hand side can be separated into two regions. If $\alpha \pF H\lesssim\beta|\Delta|$, it is negligibly small and the SC OP does not change quantitatively due to the presence of Rashba SOC.
        \revtwo{However, when $\alpha \pF H\gtrsim\beta|\Delta|$, it dominates the self-consistency equation and causes the SC OP to vanish abruptly with respect to $H/\beta$.}

		The magnetic field dependence of the SC OP in~\cref{eq:self-consistency-equation-final} can be solved in the zero temperature limit and gives
		\begin{equation}\label{eq:self-consistent-gap-equation-low-temperature-expansion}
			\frac{|\bar\beta\Delta|}{\Delta_0}
			\simeq
			\left(\frac{\Delta_0}{2\sqrt{H^2+\beta^2}}\right)^{{H^2}/{\beta^2}}
			\Theta\left(|\beta\Delta|-\alpha \pF H\right),
		\end{equation}
		with $\Theta$ the Heaviside step function being the low temperature limit of the Fermi-Dirac distribution $ \nF\left[({\alpha \pF\bar H-|\bar\beta\Delta|})/{\kB T}\right]$.
		\revtwo{This is the main result of this article. Namely, Rashba SOC induces a first-order QPT at the quantum critical point (QCP), $H_\alpha$, defined by $\alpha\pF H_\alpha=\beta\Delta(H_\alpha)$. This is a first order transition only at zero temperature, where $H_\alpha$ is equal to the critical magnetic field $H_c$. At low temperatures, the transition obtains a small broadening, since $H_\alpha<H_c$, and the SC OP vanishes rapidly in the vicinity of 
        the QCP,
        see~\cref{fig:free-energy-main}. Due to the highly non-linear magnetic field dependence of $\Delta$ in~\cref{eq:self-consistency-equation-final} close to the QCP, this abrupt change is absent in the high temperature Ginzburg-Landau limit $|\Delta|\ll \kB T$, see Appendix~\ref{app:high-temp-limit}.
        In Appendix~\ref{app:free-energy}, we furthermore derive the free energy under generalized conditions and discuss the transition in more detail.}
        
\begin{figure}
            \centering
            \includegraphics[scale=1]{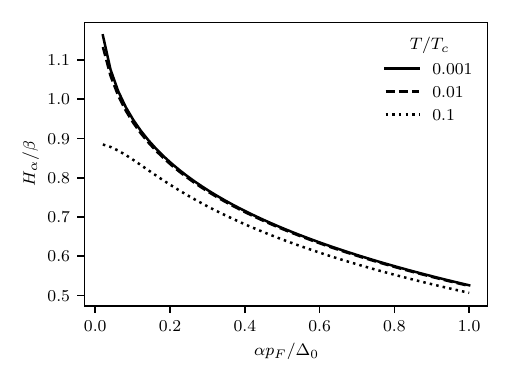}
			\caption{Dependence of $H_\alpha\alpha\pF=|\beta\Delta(H_\alpha,T)|$ on the strength of Rashba SOC for $\beta=7\Delta_0$. We notice that Rashba SOC is significant if it is one order smaller than the superconducting gap $\alpha \pF/\Delta_0\gtrsim 10^{-1}$.}
			\label{fig:critical-magnetic-field-rashba}
		\end{figure}

		For completeness, we determine the critical magnetic field due to Rashba SOC.
		We first note that~\cref{eq:self-consistent-gap-equation-low-temperature-expansion} can be approximated further by $|\Delta|/\Delta_0\sim\left(\Delta_0/2\beta\right)^{{H^2}/{\beta^2}}$.
		The critical size of the SC OP follows from $\alpha\pF H=\beta\Delta$, which together with the previous equality gives
		$\alpha\pF H/\beta=\Delta_0\left(\Delta_0/2\beta\right)^{{H^2}/{\beta^2}}$.
		We find the critical magnetic field to be given by
		$
		{H_\alpha^2}/{\beta^2}={-W[-2(\Delta_0/\alpha \pF)^2\log(\Delta_0/2\beta)]}/{2\log(\Delta_0/2\beta)},
		$
		with $W$ the Lambert $W$ function.
		We show the dependence of the critical magnetic field as a function of Rashba SOC in~\cref{fig:critical-magnetic-field-rashba}.
		We find the effect of Rashba SOC to be significant in the superconducting phase diagram for $\alpha\pF/\Delta_0\gtrsim 10^{-1}$.
		In Refs.~\cite{lu_evidence_2015,lu_full_2018} values of $\alpha \pF\sim\Delta_0$ in $\mathrm{MoS_2}$ and $\mathrm{WS_2}$ were reported.
		\revtwo{This should be sufficient for the first order QPT to be within experimental reach.}
        
\begin{figure}[t]
            \centering
            \includegraphics[scale=1]{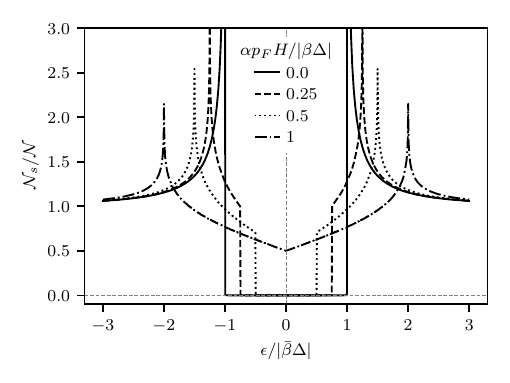}
			\caption{Density of states of the Ising SC for different values of $\alpha \pF H/|\beta\Delta|$. This describes a BCS density of states for $\alpha\pF=0$ with a gap given by $|\bar\beta\Delta|$ and becomes cone-shaped within $|\epsilon-|\bar\beta\Delta||<\alpha\pF H$ for increasing values of $\alpha\pF H/|\beta\Delta|$.
            At finite temperatures, $H_\alpha<H_c$, and the Bogoliubov Fermi surface goes through a Lifshitz transition. This renders the density of states gapless and cone-shaped for higher magnetic fields.}
			\label{fig:density-of-states}
		\end{figure}

		\revtwo{Typically, the SC OP is measured via the density of states using tunneling techniques.
        As we will show, the effective SC gap in the density of states differs from $|\Delta|$ by a factor of $\sqrt{1+H^2/\beta^2}$.
        Up to the precision of the Gorkov Green's function in~\cref{eq:1st-order-greens-function}, the density of states is given by
        \[
        \frac{\mathcal{N}_s(\epsilon)}{\mathcal{N}}
        =
        \mathrm{Re}\int_{0}^{2\pi}\frac{\mathrm d\phi}{2\pi}\,\frac{\epsilon+\alpha \pF\bar H\cos(\phi)}{\sqrt{(\epsilon+\alpha \pF\bar H\cos(\phi))^2-|\bar\beta\Delta|^2}}.
        \]
        Hence, the effective gap that is measured via the density of states is $|\bar\beta\Delta|$ rather than $|\Delta|$
        and Rashba SOC makes the density of states cone-shaped for energies $\epsilon$ satisfying $|\epsilon-|\bar\beta\Delta||<\alpha\pF H$, see~\cref{fig:density-of-states}. We want to highlight that the density of states at zero energy becomes finite once $H=H_\alpha$. This allows the population of quasiparticle states and leads to the first order QPT.
        At finite temperatures, $H_\alpha<H_c$, and the Bogoliubov Fermi surface goes through a Lifshitz transition before reaching the normal state, see~\cref{app:free-energy}. This renders the density of states gapless for $H_\alpha<H<H_c$.
        Moreover, inclusion of the second order correction in the Gorkov Green's function~\eqref{eq:1st-order-greens-function} gives rise to what is called a mirage gap at high energies $\epsilon\sim\sqrt{\beta^2+H^2}$ in the density of states~\cite{tang_magnetic_2021}.}

		\section{Discussion}
		\revtwo{In this article, we showed that the presence of Rashba SOC in Ising superconductors introduces a first order quantum phase transition at the magnetic field $\alpha\pF H_\alpha=|\beta\Delta|$. At finite temperatures this transition becomes a second order phase transition and disappears for temperatures close to $T_c$.}
        
		We find Rashba SOC to be significant to the superconducting phase diagram if $\alpha\pF/\Delta_0\gtrsim 10^{-1}$.
		Due to the heavy atom nature of TMDs, such a Rashba SOC could be induced from gating or interfacial effects in experiments.
		Experimentally, Rashba SOCs of magnitude of the SC OP have already been measured in thin-film Ising SCs~\cite{lu_evidence_2015,lu_full_2018}.
		Our findings may lead to new ways to determine the size of Rashba SOC in Ising SCs, which is currently a challenging task.
        
		The Rashba induced first order QPT should be experimentally verifiable using tunneling spectroscopy at low temperatures~\cite{kuzmanovic_tunneling_2022,dvir_spectroscopy_2018,costanzo_tunnelling_2018}.
		While different tunneling devices have different sensitivities towards either the $\Gamma$ pocket or the $\pm\mathrm{K}$ valleys, the predicted effect should be within experimental reach using tunneling devices that are mostly sensitive to the density of states in the $\pm\mathrm{K}$ valleys~\cite{kuzmanovic_tunneling_2022}.

		In this work, we considered the $s$-wave two valley model of Ising SCs and disregarded the influence of the $\Gamma$ pocket.
		This is justified for measurement devices sensitive to the valleys and for high magnetic fields---since Ising SOC is larger in the valleys.
		Furthermore, we considered a clean Ising SC and assumed tunneling between the $\Gamma$ pocket and the $\pm\mathrm{K}$ valleys to be negligible.

		In future work, one could extend this model to include the $\Gamma$ pocket~\cite{engstrom_upper_critical_field_2025}.
		Additionally, inclusion of impurities and scattering between the $\pm\mathrm{K}$ valleys and $\Gamma$ pocket would be of interest. 
        
        \revtwo{During the revision of this manuscript, we became aware of the work~\cite{wang2026} which numerically confirms the results presented in this article.}

		\section*{Acknowledgements}
		JSH would like to thank P. M. Gunnink for input on the manuscript. We acknowledge financial support by Deutsche Forschungsgemeinschaft (DFG, German Re-search Foundation)–Project-ID 443404566 - SPP 2244.
		\bibliography{bibliography-ising}
		\appendix
		\begin{widetext}
			\section{Self-consistency equation in the low temperature limit}\label{app:self-consistency-low-temperature}
			The self consistency equation from the original basis to the rotated basis is given by
			\begin{equation}
				\begin{aligned}
					\frac{\Delta}{\lambda}
					=&
					\frac{\kB T}{2}\sum_n\int\frac{\mathrm d\phi}{2\pi}
					\mathrm{Tr}\left\{f_{\ii \omega_n}\ii \sigma_y\right\}
					\rightarrow
					\frac{\kB T}{2}\sum_n\int\frac{\mathrm d\phi}{2\pi}
					\mathrm{Tr}\left\{f_{\ii \omega_n}\left(\mathcal U_\mathbf{-k}^T\ii \sigma_y\mathcal U_\mathbf k\right)\right\}.
				\end{aligned}
			\end{equation}
			Filling in the anomalous Green's function, we find the self-consistency equation~\eqref{eq:self-consistency-equation}, which reads
			\begin{equation}
				\begin{aligned}\label{eq:self-consistency-equation-app}
					\frac{1}{\N\lambda}
					=
					\pi\ii \kB T\sum_{n,\nu}&\int\frac{\mathrm d\phi}{2\pi}
					\left[
					\frac{\bar\beta^2}{\ii \sqrt{\left(\omega_n+\ii \big[\alpha\nu (\mathbf{\bar H}\times\hat{z}).\mathbf \pfnu\big]\right)^2+|\bar\beta\Delta|^2}}
					+
					\frac{\bar H^2\mathrm{sign}(\omega_n)}{\ii \omega_n-\nu\sqrt{H^2+\beta^2}}
					\right].
				\end{aligned}
			\end{equation}
			We furthermore assume $\kB T\ll\alpha \pF\lesssim \Delta_0\ll\sqrt{H^2+\beta^2}$ and notice that the sum over Matsubara frequencies of the left term can be transformed into an integral \rev{over energy}
			\begin{align}\label{eq:self-consistency}
				&\frac{1}{\mathcal N\lambda}
				=\\\nonumber
				&\sum_\nu\int\frac{\mathrm d\epsilon}{4}\int\frac{\mathrm d\phi}{2\pi}\;
				\frac{		\bar\beta^2	\tanh\left[{\left(\sqrt{\epsilon^2+|\bar\beta\Delta|^2}+\alpha\nu (\mathbf{\bar H}\times\hat{z}).\mathbf \pfnu\right)}/{	2\kB T}\right]
				}
				{
					\sqrt{\epsilon^2+|\bar\beta\Delta|^2}
				}
				+
				\pi\ii \kB T \sum_{n,\nu}
				\Bigg[
				\frac{\bar H^2\mathrm{sign}(\omega_n)}{\ii \omega_n-\nu\sqrt{H^2+\beta^2}}
				\Bigg].
			\end{align}
			In order to perform the integral in~\cref{eq:self-consistency}, we use the identity $\tanh(x)=1-2\nF(x)$. By noticing that $1/\mathcal N\lambda=\bar\beta^2/\mathcal N\lambda+\bar H^2/\mathcal N\lambda$, we may write the first line in~\cref{eq:self-consistency} as
			\begin{equation}\label{eq:energy-integral-equal-spin-sector}
				-\frac{1}{\mathcal N\lambda}
				+
				\sum_\nu\int\frac{\mathrm d\epsilon}{4}\frac{1}{\sqrt{\epsilon^2+\bar\beta|\Delta|^2}}
				-
				\int\frac{\mathrm d\epsilon}{2}\frac{\mathrm d\phi}{2\pi}\;
				\frac{\nF\left[{\left(\sqrt{\epsilon^2+|\bar\beta\Delta|^2}+\alpha\nu (\mathbf{\bar H}\times\hat{z}).\mathbf \pF\right)}/{\kB T}\right]}{\sqrt{\epsilon^2+|\bar\beta\Delta|^2}}.
			\end{equation}
			The first two terms in~\cref{eq:energy-integral-equal-spin-sector} give $\log(\Delta_0/\bar\beta|\Delta|)$.
			The integral over the Fermi-distribution on the right-hand side of~\cref{eq:energy-integral-equal-spin-sector} may be rewritten as a series expansion
			\begin{align}\label{eq:fermi-distribution-expansion}
				&\int\frac{\mathrm d\phi}{2\pi}
				\nF\left[
				{\left(
					\sqrt{\epsilon^2+|\bar\beta\Delta|^2}
					+
					\alpha\nu (\mathbf{\bar H}\times\hat{z}).\mathbf \pF
					\right)}/{
					\kB T
				}
				\right]
				\\\nonumber
				=&
				\int\frac{\mathrm d\phi}{2\pi}
				\sum_{n=1}^{\infty}
				(-1)^{n-1}\exp\left[-n\frac{\sqrt{\epsilon^2+|\bar\beta\Delta|^2}+\alpha\nu (\mathbf{\bar H}\times\hat{z}).\mathbf \pF}{\kB T}\right]
				\\\nonumber
				=&
				\sum_{n=1}^{\infty}
				(-1)^{n-1}\exp\left[-n\frac{\sqrt{\epsilon^2+|\bar\beta\Delta|^2}}{\kB T}\right]
				I_0\left(n\frac{\alpha \pF \bar H}{\kB T}\right).
			\end{align}
			For $\alpha \pF\gg \kB T$ the modified Bessel function becomes
			$
			I_0\left({n\alpha \pF\bar H}/{\kB T}\right)
			\simeq
			\sqrt{{\kB T}/{2\pi n\alpha \pF\bar H}}\exp\left({n\alpha \pF\bar H}/{\kB T}\right).
			$
			Similarly, for $|\Delta|\gg \kB T$ we may approximate the exponential in~\cref{eq:fermi-distribution-expansion} by
			$
			{\exp\left(-n{\sqrt{\epsilon^2+|\bar\beta\Delta|^2}}/{\kB T}\right)}/{\sqrt{\epsilon^2+|\bar\beta\Delta|^2}}
			\simeq
			{\exp\left(-n{(\bar\beta|\Delta|+\epsilon^2/2\bar\beta|\Delta|)/\kB T}\right)}/{\bar\beta|\Delta|}.
			$
			Using these approximations, the integral over energy in~\cref{eq:self-consistency} reads
			\begin{align}
				&\int\frac{\mathrm d\epsilon}{2}\frac{\mathrm d\phi}{2\pi}\;
				\frac{
					\nF\left[
					{\left(
						\sqrt{\epsilon^2+|\bar\beta\Delta|^2}
						+
						\alpha\nu (\mathbf{\bar H}\times\hat{z}).\mathbf \pF
						\right)}/{
						\kB T
					}
					\right]
				}
				{
					\sqrt{\epsilon^2+|\bar\beta\Delta|^2},
				}
				\\&\nonumber
				\simeq
				\sum_{n=1}^{\infty}
				\int\frac{\mathrm d\epsilon}{2}\;
				(-1)^{n-1}\frac{\exp\left[-\frac{n\epsilon^2}{2\bar\beta|\Delta|\kB T}\right]}{\bar\beta|\Delta|}
				\sqrt{\frac{\kB T}{2\pi n|\alpha|}}
				\exp\left[\frac{n(\alpha \pF\bar H-\bar\beta|\Delta|)}{\kB T}\right]
				\\&\nonumber
				=
				\sum_{n=1}^{\infty}
				\frac{(-1)^{n-1}}{2n}
				\frac{\kB T}{\sqrt{\bar\beta|\Delta|\alpha \pF\bar H}}
				\exp\left[\frac{n(\alpha \pF\bar H-\bar\beta|\Delta|)}{\kB T}\right]
				\\&\nonumber
				=
				\frac{1}{2}
				\frac{\kB T}{\sqrt{\bar\beta|\Delta|\alpha \pF\bar H}}
				\log\left(1+\exp\left[\frac{\alpha \pF\bar H-\bar\beta|\Delta|}{\kB T}\right]\right).
			\end{align}

			Moreover, the sums over Matsubara frequencies on the right-hand hand side of~\cref{eq:self-consistency} are known to give polygamma functions.
			To obtain the final expression for the self-consistency equation, we reinterpret the BCS energy cutoff $\wD$ in terms of $T_c$ and $\Delta_0$.
			To achieve this, we use that $1/\mathcal N \lambda=\log(2\wD/|\Delta_0|)=\sum_{n=0}^{\hbar\wD/2\pi \kB T_c}{(n+1/2)^{-1}}$ at zero field.
			Hence, the self-consistency equation at low temperatures $\Delta\gg \kB T$ simplifies to
			\begin{align}\label{eq:self-consistency-equation-final-app-1}
				\log\left(\frac{|\Delta_0|}{\bar\beta|\Delta|}\right)
				=&
				\sum_\nu
				\frac{1}{2}
				\frac{\kB T}{\sqrt{\bar\beta|\Delta|\alpha \pF\bar H}}
				\log\left(1+\exp\left[\frac{\alpha \pF\bar H-\bar\beta|\Delta|}{\kB T}\right]\right)
				+
				\frac{ H^2}{\beta^2}
				\Bigg[
				\log\left(\frac{T}{T_c}\right)
				-
				\psi\left(\frac{1}{2}\right)
				+
				\log\left(\frac{\sqrt{H^2+\beta^2}}{2\pi \kB T}\right)
				\Bigg].
			\end{align}
			By using the BCS relation $\Delta_0=\pi \kB T/\exp(\gamma)$ and $\psi(1/2)=-2\log(2)-\gamma$,~\cref{eq:self-consistency-equation-final-app-1} reads
			\begin{align}\label{eq:self-consistency-equation-final-app}
				\log\left(\frac{|\Delta_0|}{\bar\beta|\Delta|}\right)
				=&
				\frac{\kB T}{\sqrt{\bar\beta|\Delta|\alpha \pF\bar H}}
				\log\left(1+\exp\left[\frac{\alpha \pF\bar H-\bar\beta|\Delta|}{\kB T}\right]\right)
				+
				\frac{ H^2}{\beta^2}
				\log\left(\frac{2\sqrt{H^2+\beta^2}}{\Delta_0}\right).
			\end{align}

			\section{High temperature Ginzburg-Landau limit}
            \label{app:high-temp-limit}
			To determine the high temperature limit, $(T_c-T)/T\ll 1$, we expand the self-consistency equation~\eqref{eq:self-consistency-equation-app} up to second order in $\bar\beta|\Delta|$ and $\alpha \pF$ and ignore the contribution from the helical state $\vF q\rightarrow0$.
			After taking the angular integral, the self-consistency equation~\eqref{eq:self-consistency-equation-app} reads
			\begin{equation}
				\begin{aligned}
					\frac{1}{\N\lambda}
					=
					\pi\ii \kB T\sum_{n,\nu}\;
					\bar\beta^2 \mathrm{sign}(\omega_n)
					&\left[
					\frac{1}{\ii {\omega_n}}
					-\frac{\alpha^2p_{F}^2\bar H^2+|\bar\beta\Delta|^2}{2\ii \omega_n^3}
					\right]
					\\
					+
					{\bar H^2\mathrm{sign}(\omega_n)}
					&\left[\frac{1}{\ii \omega_n-\nu\sqrt{H^2+\beta^2}}\right].
				\end{aligned}
			\end{equation}
			The sums over Matsubara frequencies are known to give polygamma functions and the above self-consistency equation becomes
			\begin{equation}
				\begin{aligned}
					\log\left(\frac{T}{T_c}\right)
					=
					\bar H^2
					&\left[
					\psi\left(\frac{1}{2}\right)
					-
					\mathrm{Re}
					\psi\left(\frac{1}{2}-\ii \frac{\sqrt{H^2+\beta^2}}{2\pi \kB T}\right)
					\right]
					-
					\bar\beta^2
					\left[
					\frac{7\zeta(3)}{2(2\pi \kB T)^2}\left(\alpha^2p_{F}^2\bar H^2+|\bar\beta\Delta|^2\right)
					\right].
				\end{aligned}
			\end{equation}
			Since the Ising SOC is typically much larger than the critical temperature $\sqrt{\beta^2+H^2}\gg \kB T$ the digamma function is well approximated by a logarithm
			$
			\mathrm{Re}\psi\left({1}/{2}-\ii {\sqrt{H^2+\beta^2}/}{2\pi \kB T}\right)
			\rightarrow
			\log\left({\sqrt{H^2+\beta^2}}/{2\pi \kB T}\right).
			$
			Thus, the above is well approximated by
			\begin{equation}
				\begin{aligned}
					\log\left(\frac{T}{T_c}\right)
					=
					\bar H^2
					\left[
					\psi\left(\frac{1}{2}\right)
					-
					\log\left(\frac{\sqrt{H^2+\beta^2}}{2\pi \kB T}\right)
					\right]
					-\bar\beta^2
					&\left[
					\frac{7\zeta(3)}{2(2\pi \kB T)^2}\left(\alpha^2p_{F}^2\bar H^2+|\bar\beta\Delta|^2\right)
					\right]
					.
				\end{aligned}
			\end{equation}
			The above can be written more compactly as
			\begin{equation}
				\begin{aligned}
					\frac{7\zeta(3)}{2(2\pi \kB T)^2}\left(|\bar\beta\Delta|^2+\alpha^2p_{F}^2\bar H^2\right)
					=
					\log\left(\frac{T_c}{T}\right)
					-
					\frac{H^2}{\beta^2}
					\log\left(\frac{2\sqrt{H^2+\beta^2}}{\Delta_0}\right).
				\end{aligned}
			\end{equation}
			Hence, we see that the SC OP does not undergo a first order phase transition at $\beta|\Delta|=\alpha \pF H$ for temperatures close to $T_c$.
            \begin{figure}
                 \centering
                \includegraphics[scale=1]{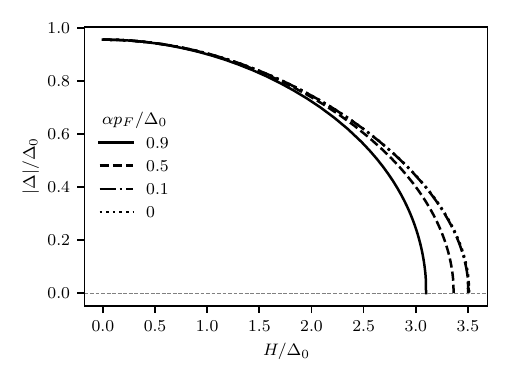}
                \caption{Numerical solution of the self-consistency equation from~\cref{eq:sce_numeric} for $|\bar\beta\Delta|$ as a function of the external magnetic field $H$ for different values of Rashba SOC and $\beta=7\Delta_0$ at $T=0.5T_c$.}
                \label{fig:gap_GL}
            \end{figure}
            
            \section{Free energy and phase diagram}
            \label{app:free-energy}
            We compute the free energy difference $\F=\F_S-\F_N$ with respect to the normal state. Within the functional-integral treatment of BCS mean-field theory, the free energy of the superconductor reads
            \begin{align}
                \F_S[\Delta,\Delta^*] = -\frac12 \sum_{\mathbf k, \mathbf k'} \sum_{s_i,s_i'} \Delta^\ast_{s_1's_2'}(\mathbf k') W^{-1}_{s_1's_2',s_1s_2}(\mathbf k ,\mathbf k')\Delta_{s_1s_2}(\mathbf k) - \frac T2\sum_{\mathbf k,\omega_n}\log(\det(-\check\G_{\ii\omega_n}^{-1})).
            \end{align}
            This is a standard result which can be found for example in A.~Altland and B.~D.~Simons, \emph{Condensed Matter Field Theory}, 2nd ed., Cambridge University Press, 2010, chapter on superconductivity. In this work, $W$ denotes the attractive pairing interaction in the spin-singlet Cooper channel, constant within the shell $|\xi_{\mathbf k}|<\wD$. Hence, the first term reduces to $|\Delta|^2/V$ with constant $V=\N\lambda>0$ and the free energy reads
            \begin{align}
                \F = \frac{|\Delta|^2}{V} - \frac T2\sum_{\mathbf k,n}\log(\frac{\det\check\G_{\ii\omega_n}^{-1}}{\det\check\G_{\ii\omega_n,0}^{-1}})
            \end{align}
            with $\check\G_{\ii\omega_n,0}^{-1}=\check\G_{\ii\omega_n}^{-1}(\Delta\to0)$, the normal state Green's function. We identify
            \begin{align}
                \label{eq:ln_det}
                \log(\frac{\det\check\G_{\ii\omega_n}^{-1}}{\det\check\G_{\ii\omega_n,0}^{-1}}) = \sum_\nu\log(\frac{x_\nu^2 + \Omega_\nu^2}{x_\nu^2 + \Omega_{\nu,0}^2}) + \log(\det(\mathds 1-\check\G^0_{\ii\omega_n,-}\check\Sigma_-)),
            \end{align}
            where we have introduced 
            \[
                \Omega_\nu\equiv\sqrt{(\omega_n - \ii\nu a)^2+|\bar\beta\Delta|^2}, \quad a\equiv\alpha(\bar{\mathbf{H}}\times\hat z).\mathbf p_{\rm F}, \quad x_\nu\equiv\frac12[\epsilon_\nu(\mathbf k)+\epsilon_\nu(-\mathbf k)] = \frac{\mathbf p^2}{2m^*} - \mu + \nu B + \mathcal O(\alpha^2\pF^2/B)
            \]
            and $B\equiv\sqrt{\beta^2+H^2}$ as well as the self-energy due to interband coupling $\check\Sigma_-=\check D_{-+}\check\G^0_{\ii\omega_n,+}\check D_{+-}$ with
            \begin{align}
                \check D_{\nu\bar\nu} = \begin{pmatrix} 0 & \Delta (\mathcal U^\dagger_{\mathbf k}\,\ii\sigma_y\,\mathcal U^*_{-\mathbf k})_{\nu\bar\nu}\\ \Delta^* (\mathcal U^T_{-\mathbf k}\,\ii\sigma_y\,\mathcal U_{\mathbf k})_{\bar\nu\nu} & 0 \end{pmatrix}.
            \end{align}
            The first term and second term in~\eqref{eq:ln_det} respectively denote the intraband and interband contribution to the free energy. Up to second order in interband coupling $|\bar H\Delta|$ and with $\log(\det(\mathds 1-\check\G^0_{\ii\omega_n,-}\check\Sigma_-))\approx-\tr(\check\G^0_{\ii\omega_n,-}\check\Sigma_-) + \mathcal O(|\bar H\Delta|^4)$, we find for the second term
            \begin{align}
                \sum_{\mathbf k} \tr(\check\G^0_{\ii\omega_n,-}\check\Sigma_-)\approx\sum_\eta\int_0^{2\pi}\frac{{\mathrm d}\phi}{2\pi}\sum_\nu\int dx_\nu\,\N_\nu\tr_\nu(\check M) = -\pi\N\mathrm{sgn}(\omega_n)\frac{|\bar H\Delta|^2}{2}\sum_\nu\frac{1}{\omega_n+\ii\nu B}
            \end{align}
            within the quasi-classical approximation, where $\N_+\simeq\N_-\simeq\N/4$ -- the density of states is approximately independent of valley and band. To compute the first term, we use
            \begin{align}
                \partial_d\int {\mathrm d}x_\nu\,\log(\frac{x_\nu^2 + \Omega_\nu^2}{x_\nu^2 + \Omega_{\nu,0}^2}) = \int {\mathrm d}x_\nu\,\frac{\partial_d\Omega_\nu^2}{x_\nu^2 + \Omega_\nu^2} = \int \frac{{\mathrm d} x_\nu}{x_\nu^2 + \Omega_\nu^2} = \frac{\pi}{\Omega_\nu} = 2\pi\partial_d\Omega_\nu
            \end{align}
            with $d=|\bar\beta\Delta|^2$ and $\Re(\Omega_\nu)>0, \forall d$. Both sides vanish for $d=0$, where $\Omega_\nu=\Omega_{\nu,0}$ and integration over $d$ leads to
            \begin{align}
                \int {\mathrm d}x_\nu\,\log(\frac{x_\nu^2 + \Omega_\nu^2}{x_\nu^2 + \Omega_{\nu,0}^2}) = 2\pi\int_0^d {\mathrm d}d'\,\partial_d\Omega_\nu(d') = 2\pi (\Omega_\nu(d)-\Omega_{\nu,0}).
            \end{align}
            Since $\Omega_\nu(-\omega_n)=\Omega_\nu(\omega_n)^*$, $\Omega_{\nu,0}(-\omega_n)=\Omega_{\nu,0}(\omega_n)^*$ and $\mathrm{sgn}(-\omega_n)/(-\omega_n+\ii\nu B)=[\mathrm{sgn}(\omega_n)/(\omega_n+\ii\nu B)]^*$, the terms with $\pm\omega_n$ are complex conjugates of each other. The sum over $n\in\mathbb Z$ therefore equals $2\Re$ of the sum over $n\ge0$. Taking the real part furthermore takes $\Omega_{\nu,0}\to\omega_n$ and the free energy follows as
            \begin{align}
                \F = \frac{|\Delta|^2}{V} - 4\pi T\sum_{n\ge0}\sum_\nu\int\frac{{\mathrm d}\phi}{2\pi}\N_\nu\Re\big(\Omega_\nu-\omega_n\big) - 2\pi T|\bar H\Delta|^2\sum_{n\ge0}\sum_\nu\int\frac{{\mathrm d}\phi}{2\pi}\N_\nu\Re\frac1{\omega_n+\ii\nu B},
            \end{align}
            where we have performed the sum over valleys in both terms. After some algebraic steps and a replacement of the interaction strength with the critical temperature, we arrive at the equivalent expression
            \begin{align}
                \frac\F\N &= \frac{|\Delta|^2}{2}\log(\frac{T}{T_c}) + \frac{|\bar H\Delta|^2}{2} \Re\left[\psi\left(\frac12+\frac{\ii B}{2\pi T}\right) - \psi\left(\frac12\right)\right] - \frac{|\bar\beta\Delta|^2}{2}\left[ \frac12 + \log(\frac T{T_c}) + \log(\frac{\Delta_0}{|\bar\beta\Delta|}) \right] \notag\\
                &\phantom{=}- T\sum_{\sigma=\pm}\int\frac{{\mathrm d}\phi}{2\pi}\int_0^\infty {\mathrm d}\xi\,\left[\log(1+\exp(-\frac{E+\sigma a}{T})) - \log(1+\exp(-\frac{\xi+\sigma a}{T})) \right]
            \end{align}
            with $E=\sqrt{\xi^2+|\bar\beta\Delta|^2}$ and $\Delta_0=\pi e^{-\gamma}T_c$. For ISCs, it is safe to assume $T\ll B$ as the Ising SOC strength $\beta>T_c$. Hence, $B$ is much larger than all temperatures admissible in the superconducting state and the free energy reduces to
            \begin{align}
                \frac\F\N &= \frac{|\bar H\Delta|^2}{2}\log(\frac{2B}{\Delta_0}) - \frac{|\bar\beta\Delta|^2}{2}\left[ \frac12 + \log(\frac{\Delta_0}{|\bar\beta\Delta|}) \right] \notag\\
                \label{eq:free_energy}
                &\phantom{=} - T\sum_\sigma\int\frac{{\mathrm d}\phi}{2\pi}\int_0^\infty {\mathrm d}\xi\,\left[\log(1+\exp(-\frac{E+\sigma a}{T})) - \log(1+\exp(-\frac{\xi+\sigma a}{T})) \right].
            \end{align}
            We show the free energy and the influence of Rashba SOC on the free energy close to the transition in~\cref{fig:free-energy-alpha}.
            \begin{figure}[htb]
                \centering
                \begin{subfigure}[b]{.49\textwidth}
                    \centering
                    \includegraphics[scale=1]{free_en_ap_0p1_T_0p01.pdf}
                \end{subfigure}
                \begin{subfigure}[b]{.49\textwidth}
                    \centering
                    \includegraphics[scale=1]{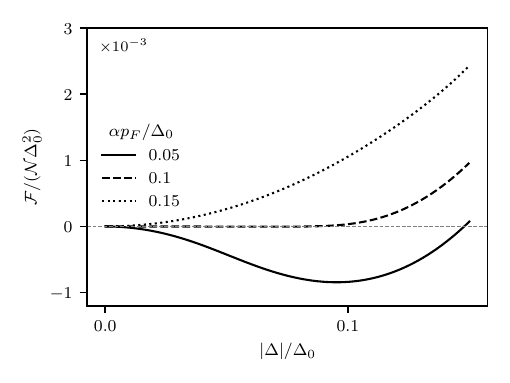}
                \end{subfigure}
                \caption{(left) Free energy as a function of the normalized gap for $\beta=7\Delta_0$, $\alpha\pF=0.1\Delta_0$, $T/T_c=10^{-2}$ and different values of $H$. Close to the transition, at $H\approx6.65\Delta_0$, the shape of the free energy depends on the value of $\alpha\pF$ as shown on the right.}
                \label{fig:free-energy-alpha}
            \end{figure}
            The self-consistency follows from the stationarity condition $\frac{\partial\F}{\partial\Delta^*} = \frac{\partial|\bar\beta\Delta|}{\partial\Delta^*} \frac{\partial\F}{\partial|\bar\beta\Delta|} = 0$ and reads
            \begin{align}
                \label{eq:sce_numeric}
                \log(\frac{\Delta_0}{|\bar\beta\Delta|}) &= \frac{H^2}{\beta^2}\log(\frac{2B}{\Delta_0}) + \sum_\sigma\int\frac{{\mathrm d}\phi}{2\pi}\int {\mathrm d}\xi\,\frac{n_{\rm F}(E+\sigma a)}{E}
            \end{align}
            which in the limit $T\ll\alpha\pF\ll|\bar\beta\Delta|$ reduces to~\eqref{eq:self-consistency-equation-final-app}. 
            At $T=0$, $n_{\rm F}(\epsilon)\to\Theta(-\epsilon)$ and the last term in~\cref{eq:sce_numeric} becomes
            \begin{align}
                K\equiv\sum_\sigma\int\frac{{\mathrm d}\phi}{2\pi}\int_0^\infty\frac{{\mathrm d}\xi}{E}\Theta(-(E+\sigma a)) = \int\frac{{\mathrm d}\phi}{2\pi}\text{arccosh}\left(x|\cos(\phi)|\right)\Theta(x|\cos(\phi)|-1), \qquad x\equiv 
                \frac{H\alpha\pF}{|\beta\Delta|}.
            \end{align}
            For $H\alpha\pF\le|\beta\Delta|$, $K=0$ whereas for $H\alpha\pF\ge|\beta\Delta|$ the angular integral gives $K=\log(H\alpha\pF/|\beta\Delta|)$. Inserting this into the self-consistency equation, $\Delta$ drops out and one obtains $\Delta_0 \left(2B/\Delta_0\right)^{-H^2/\beta^2} = \bar H\alpha\pF$ which defines $H_\alpha$. To show that it coincides with the critical field at $T=0$, we note that integrating the self-consistency equation for $H\alpha\pF>|\beta\Delta|$ leads to
            \begin{align}
                \frac\F\N\propto\frac{|\beta\Delta|^2}2\Big[\log(\frac{H\alpha\pF}{\Delta_0}) + \frac{H^2}{\beta^2}\log(\frac{2B}{\Delta_0})\Big].
            \end{align}
            In this regime, $K$ exactly cancels $-\log(|\beta\Delta|)$ and no quartic term stabilizes the SC OP. The first-order nature of the transition becomes evident from the ground-state energy $f(H)\equiv\min_\Delta\F$,
            \begin{align}
                \frac{f(H)}\N = -\frac14\Big[ \Delta_0^2\left(2B/\Delta_0\right)^{-2H^2/\beta^2} - \bar H^2\alpha^2\pF^2\Big] <0, \quad H<H_\alpha
            \end{align}
            whereas $f=0$ for $H\ge H_\alpha$. Since $f$ is continuous at $H_\alpha$, $\partial_Hf(H_\alpha^-)>0$ and $\partial_Hf(H_\alpha^+)=0$, the magnetization jumps at $H_c=H_\alpha$ and the transition is of first order. 
            This effect can also be understood as follows. On the cylinder $M=\mathbb R\times S^1 \ni(\xi,\phi)$, the quasiparticle energy reads
            \begin{align}
                E_\sigma(\xi,\phi) = \sqrt{\xi^2+|\beta\Delta|^2} + \sigma H\alpha\pF\cos(\phi).
            \end{align}
            For $H\alpha\pF<|\beta\Delta|$, $E_\sigma$ is strictly positive and the spectrum remains fully gapped. At $H\alpha\pF=|\beta\Delta|$, the minimum of $E_\sigma$, located at $\xi=0$ and $\sigma\cos\phi=-1$, touches zero. Upon crossing this point, $\Omega_\sigma=\{(\xi,\phi)\in M:\ E_\sigma(\xi,\phi)<0\}$ is non-empty and its boundary $\partial\Omega_\sigma=\{E_\sigma=0\}$ forms a Bogoliubov Fermi surface. More explicitly, $\Omega_\sigma$ changes from the empty set to a disk and its Euler characteristic jumps from zero to one for each $\sigma$. Such a change in topology of the Bogoliubov Fermi surface is called a Lifshitz transition~\cite{Lifshitz_1960,PhysRevLett.118.127001}. Further increasing $H\alpha\pF/|\beta\Delta|$ enlarges the region $\Omega_\sigma$ and the SC OP is suppressed by a finite density of states at the Fermi energy. As shown above, this population exactly compensates the pairing logarithm at $T=0$, and the superconducting state is not stabilized once the Bogoliubov Fermi surface appears.
            
            In~\cref{fig:phase_diagram}, we plot the normalized free energy landscape as a function of the normalized gap and for different slices of magnetic field strengths at $T=10^{-2}T_c$ on the left. On the right, we plot the phase diagram -- the QCPs are marked by red stars at $T=0$.
            \begin{figure}[htb]
                \centering
                \begin{subfigure}[b]{.49\textwidth}
                    \centering
                    \includegraphics[scale=1]{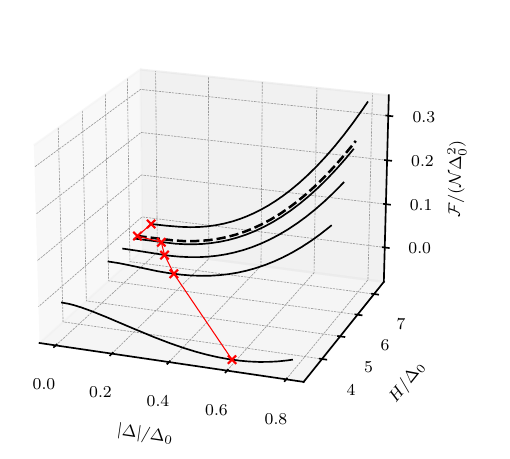}
                \end{subfigure}
                \hfill
                \begin{subfigure}[b]{.49\textwidth}
                    \centering
                    \includegraphics[scale=1]{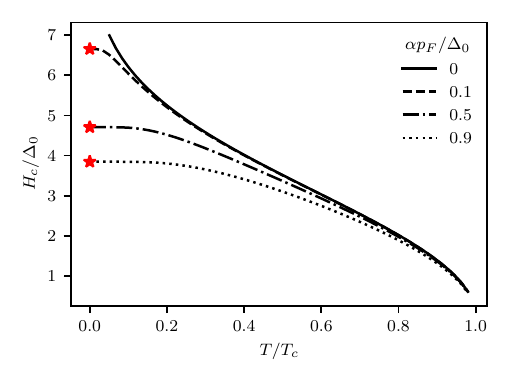}
                \end{subfigure}
                \caption{(left) The free energy landscape through slices $H/H_c=\{0.5,0.8,0.9,0.98,1,1.1\}$ for $\beta=7\Delta_0$, $\alpha\pF=0.1\Delta_0$ and $T/T_c=10^{-2}$. Indicated by the red crosses are the minima of $\F/\N$. The dashed line marks the transition to the normal state. (right) The phase diagram for $\beta=7\Delta_0$ and $\alpha\pF$ as indicated in the legend. The red stars at $T=0$ indicate the quantum QCPs.}
                \label{fig:phase_diagram}
            \end{figure}
		\end{widetext}
\end{document}